\documentclass[journal]{IEEEtran}
\IEEEoverridecommandlockouts
\usepackage{cite}
\usepackage[english]{babel}
\usepackage[T1]{fontenc}
\usepackage[utf8]{inputenc}
\usepackage{booktabs}
\usepackage{multirow}
\usepackage{multicol}
\usepackage{graphicx}
\usepackage{amsmath}
\usepackage{lipsum}
\usepackage{float}
\usepackage{geometry}
\usepackage{balance}
\usepackage{xcolor}

\usepackage[subtle]{savetrees}
\usepackage[figurename=Fig.]{caption}
\ifCLASSINFOpdf
\else
\fi
\begin{document}
\title{Physics-Informed Neural Networks for Minimising Worst-Case Violations in DC Optimal Power Flow}
\author{Rahul~Nellikkath, Spyros~Chatzivasileiadis
 \thanks{Rahul Nellikkath, and Spyros Chatzivasileiadis are with the Center for Electric Power and Energy, Technical University of Denmark, 2800 Kgs. Lyngby, Denmark. e-mail: \{rnelli, spchatz\}@elektro.dtu.dk.}
}

\markboth{}%
{Shell \MakeLowercase{\textit{et al.}}: Physics-Informed Neural Network to Minimise the Worst-Case Constraint Violations in DC Optimal Power Flow Predictions}
\maketitle
\begin{abstract}
Physics-informed neural networks exploit the existing models of the underlying physical systems to generate higher accuracy results with fewer data. Such approaches can help drastically reduce the computation time and generate a good estimate of computationally intensive processes in power systems, such as dynamic security assessment or optimal power flow. Combined with the extraction of worst-case guarantees for the neural network performance, such neural networks can be applied in safety-critical applications in power systems and build a high level of trust among power system operators. This paper takes the first step and applies, for the first time to our knowledge, Physics-Informed Neural Networks with Worst-Case Guarantees for the DC Optimal Power Flow problem. We look for guarantees related to (i) maximum constraint violations, (ii) maximum distance between predicted and optimal decision variables, and (iii) maximum sub-optimality in the entire input domain. In a range of PGLib-OPF networks, we demonstrate how physics-informed neural networks can be supplied with worst-case guarantees and how they can lead to reduced worst-case violations compared with conventional neural networks.

\end{abstract}
\section{Introduction}
Optimal Power Flow (OPF) is a tool that is increasingly used by power system operators, electricity markets, and the rest of the power system industry both for planning and real-time operation. Countless instances of OPF need to be solved when it comes to assessing uncertain scenarios, identifying transmission investments, finding optimal control setpoints, or determining the electricity market clearing. However, the accurate representation of the AC Power Flow equations in the OPF problem renders the problem non-linear and non-convex \cite{Convex}, which usually results in significant challenges related to convergence and long computing times. As a substitute, the DC-OPF approximation is often used to reduce the complexity and improve speed\cite{DCOPF}. However, even with DC-OPF, solving for multiple scenarios in a short period could be challenging. This has led to the development of various neural network (NN) architectures for predicting OPF problems. \cite{MLDCOPF} \cite{recent}. 
However, these machine learning algorithms require a considerable amount of good quality datasets to train a neural network with reasonable accuracy, while we especially need OPF results that cover both normal and abnormal situations; this is often challenging to generate. As a result, researchers have investigated different NN training approaches to reduce the dependency on the training dataset size and improve accuracy. A convex relaxation technique to only focus on the regions closer to the security boundary was proposed in \cite{Database}.  A method to incorporate adversarial examples into the training set to improve performance was introduced in \cite{Andreas_verify}. Furthermore, an input-convex NN was introduced in \cite{Convex_DCOPF} to improve its performance in unseen data points since the underlying DC-OPF problem is convex. A NN training framework to match both prediction and the sensitivity of the  OPF problem was also proposed in \cite{Sensitivity} to improve data efficiency and convergence.

However, these proposed algorithms could be susceptible to outliers. Considering OPF is often used for safety-critical applications, outliers leading to massive system constraint violations such as line, voltage, or generator limits can have a catastrophic effect on system security. It is not easy to eliminate constraint violations when using NN to estimate the OPF result. To overcome this shortcoming, the generation limits can be enforced on the NN prediction, as Ref.~\cite{Fast_ACOPF} \cite{dc3} proposed for the AC-OPF problem. However, this could lead to sub-optimality. So, a few methods have also suggested penalizing the constraint violations by adding them directly in the loss terms\cite{deepopf}. A few approaches have also used the Lagrangian dual scheme for DNN training, such as \cite{DualDNN}, and \cite{Pred_AC_OPF} for AC-OPF. Even then, these studies are trained to minimize the average prediction errors, and so far, none of the proposed machine learning algorithms have supplied any worst-case performance guarantees.     

Here, we propose a physics-informed neural network (PINN) algorithm to predict the DC-OPF problem solutions, which will reduce the dependency on the training data set, and improve the worst-case guarantees while improving optimality. PINNs for power system applications is a NN training architecture that uses the underlying physical laws governing the power system model to improve prediction accuracy \cite{misyris}. By doing that, the NN training is no longer dependent only on the size and quality of the training dataset. Instead, the training procedure can determine the NN optimal parameters based on the actual equations that the NN aims to emulate.

Our contributions in this paper are:
\begin{enumerate}
    \item For the first time, we propose a PINN architecture which utilizes KKT conditions to predict the DC-OPF problem solutions. The work in this paper is seen as the first necessary step to identify opportunities and challenges towards the application of PINN on AC-OPF problems.
    \item Extending our previous work, presented in \cite{Andreas}, we develop approaches to determine the worst-case violations of PINN, and propose ways to reduce them.
\end{enumerate}

This paper is structured as follows: Section II describes the optimal power flow problem, introduces the architecture of the physics-informed neural network, and discusses the MILP algorithm used to quantify the worst-case guarantees. Section III presents simulation results demonstrating the performance of PINN. Section IV discusses the possible opportunities to improve the system performance, and concludes. 

\section{Methodology}
\subsection{DC - Optimal Power Flow}
DC optimal power flow (DC-OPF) is a simplified linear approximation of the AC-OPF problem. A DC-OPF problem for generation cost minimization in an $N_{bus}$ system with $N_g$ number of generators and $N_{d}$ number of loads can be represented as follows:
\begin{equation}
    \mathrm{min} \  \mathbf{c^T\  P_{g}}
    \label{Obj}
\end{equation}
\begin{equation}
    \sum_{i=1}^{N_g} P_{g,i} - \sum_{i=1}^{N_d} P_{d,i} = 0
\label{ConLoadBal}
\end{equation}
\begin{equation}
    \mathbf{P_{g}^{min}} \leq \mathbf{P_{g}} \leq \mathbf{P_{g}^{max}}
    \label{ConGenLim}
\end{equation}
\begin{equation}
\lvert \mathbf{PTDF(P_{g} - P_{d})} \lvert \leq \mathbf{P_{l}^{max}}
\label{ConLineLim}
\end{equation}
where vector $\mathbf{c^T}$ is the linear cost term of each generator, vector $\mathbf{P_{g}}$ is the generator active power output and $\mathbf{P_d}$ is the active power demand. The minimum and maximum active power generation limit are denoted by $\mathbf{P_{g}^{min}}$ and $\mathbf{P_{g}^{max}}$ respectively, and $\mathbf{P_{l}^{max}}$ represents the line flow limit. $\mathbf{PTDF}$ is the power transfer distribution factors (for more details, see \cite{LecturenoteOPF}). 

The generation and line flow limits are guaranteed by \eqref{ConGenLim} and \eqref{ConLineLim}, and \eqref{ConLoadBal} ensures load balance in the system. The corresponding Karush–Kuhn–Tucker (KKT) conditions for the DC-OPF can be formulated as follows:
\begin{align}
\mathbf{c} + \lambda + \mathbf{\overline{\mu}_g} - \mathbf{\underline{\mu}_g} + \mathbf{\overline{\mu}_{l}PTDF} - \mathbf{\underline{\mu}_{l}PTDF}  &= 0 \label{Stat}\\
{\overline{\mu}_{g,i}[P_{g,i}^{max} - P_{g,i}]} &= 0 \label{ComSlak1}\\
{\underline{\mu}_{g,i}[P_{g,i} - P_{g,i}^{min}]} &= 0\\
\overline{\mu}_{l,j} \left[ \mathbf{PTDF_j(P_{g} - P_{d})} - {P_{l,j}^{max}} \right] &= 0\\
\underline{\mu}_{l,j} \left[ \mathbf{-PTDF_j(P_{g} - P_{d})} - {P_{l,j}^{max}} \right] &= 0    \label{ComSlak4}\\
\overline{\mu}_g,\underline{\mu}_g,\overline{\mu}_{l},\underline{\mu}_{l}& \geq 0\\
    \eqref{ConLoadBal} - \eqref{ConLineLim} \label{dual}
\end{align}
where $\lambda$, $\mu_{g}$ and $\mu_{l}$ denote the dual variables for \eqref{ConLoadBal} - \eqref{ConLineLim} respectively. The stationarity condition is given in \eqref{Stat}, and the complementary slackness conditions are described in \eqref{ComSlak1} - \eqref{ComSlak4}. These KKT conditions are necessary and sufficient for optimality in the DC-OPF problem \cite{boyd}, given the DC-OPF problem is feasible.  

\subsection{Physics-Informed Neural Network}\label{SecPINN}
\begin{figure}[htbp]
\centerline{\includegraphics[scale=.38]{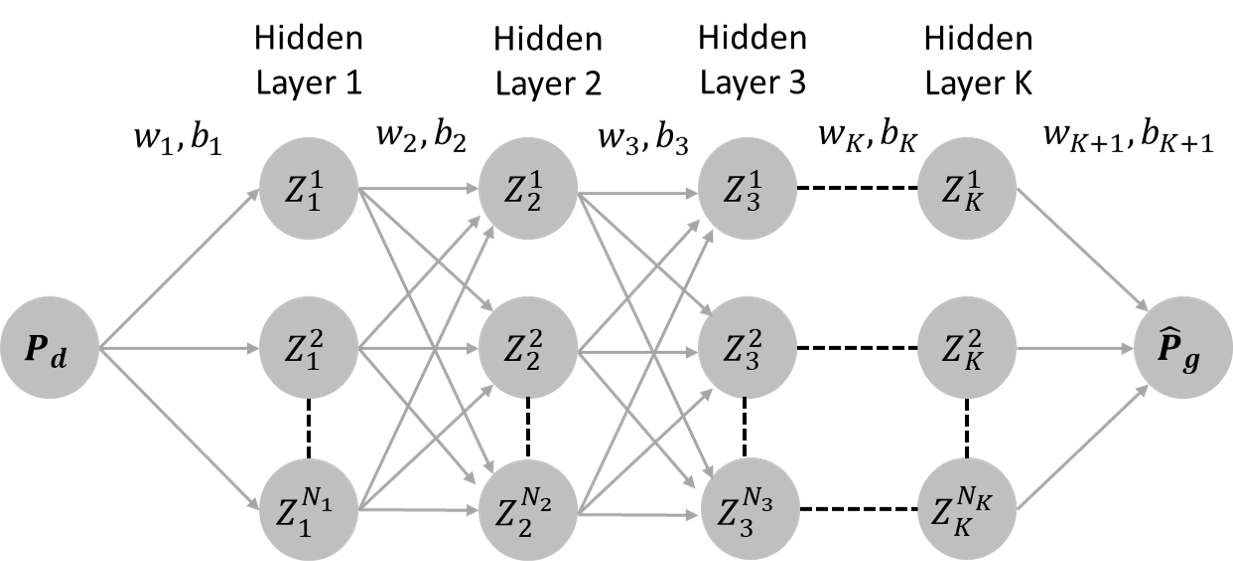}}
\caption{Illustration of the neural network architecture to predict the optimal generation outputs $\mathbf{\hat{P}_g}$ using the active power demand $\mathbf{P_d}$ as input: There are K hidden layers in the NN with $N_k$ neurons each. Where k = 1, ...,K.}
\label{NN_basic}
\end{figure}

This section introduces the PINN architecture used for predicting the DC-OPF optimal $\mathbf{P_g}$ setpoints, given active power demand $\mathbf{P_d}$ as the input. A neural network is the group of interconnected nodes connecting the input and the output layers, as shown in Fig. \ref{NN_basic}. There are $K$ number of hidden layers in the NN with $N_k$ number of neurons in the hidden layer $k$. Each neuron in the NN has a nonlinear activation function linked with them, and the edges connecting the neurons have a weight $\mathbf w$ and a bias $\mathbf b$ associated with them.

The output of each layer in the neural network can be denoted as follows:
\begin{equation}
    Z_{k+1} = \pi(w_{k+1}Z_k+b_{k+1})
\end{equation}
where $Z_{k+1}$ is the output of layer $k+1$, $w_{k+1}$ and $b_{k+1}$ are the weights and biases connecting layer $k$ and $k+1$. $\pi$ is the nonlinear activation function. As in most recent literature, in this work we chose the ReLU as the non-linear activation function, as it is observed to accelerate the NN training \cite{glorot}. The ReLU activation function will return the input if the input is positive and return zero if the input is negative. The ReLU activation function can be formulated as follows:
\begin{align}
    \hat Z_{k+1} &= w_{k+1}Z_{k}+b_{k+1}\label{NN1}\\
    Z_{k+1} &= max( \hat Z_{k+1},0)\label{Relu}
\end{align}
When we use a NN to predict the optimal generator setpoints, these weights and biases are trained to predict the generation values of the optimal setpoint for the DC-OPF problem. 

\begin{figure}[htbp]
\centerline{\includegraphics[scale=.38]{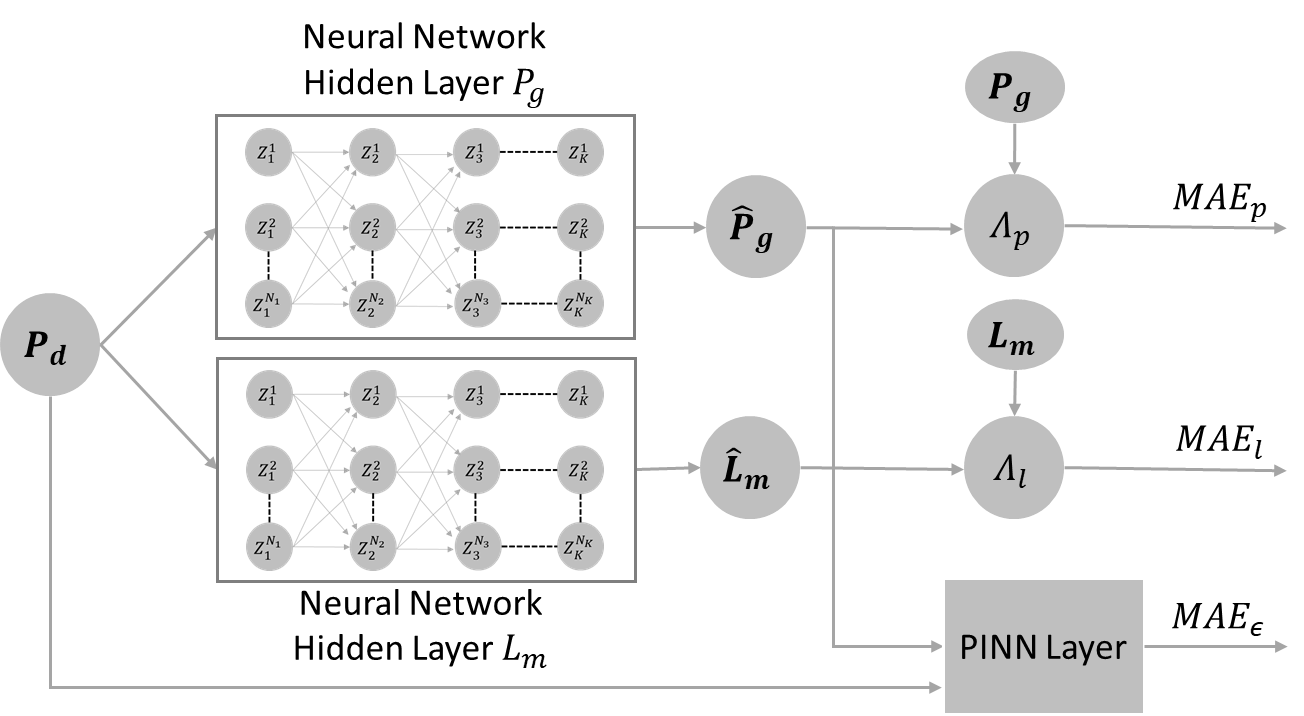}}
\caption{Illustration of the physics-informed neural network architecture to predict the optimal generation outputs $\mathbf{\hat P_g}$ and dual variables $\mathbf{\hat{L}_m}$ using the active power demand $\mathbf{P_d}$ as input. Hidden layers used for predicting $\mathbf{\hat{P}_g}$ and $\mathbf{\hat{L}_m}$ are separate and not connected to each other. During training, the NN weights (W) and biases (b) are adjusted according to loss function \eqref{MAE}, which minimizes the mean absolute errors $MAE_p$, $MAE_l$ and $MAE_{\epsilon}$}
\label{PINN_Fig}
\end{figure}

In a physics-informed neural network, the physical equations governing the problem will be incorporated into the NN loss function (see \cite{raissi_PINN}, and our previous work \cite{misyris} for power systems applications) . In the case of a DC-OPF problem, the KKT conditions given in \eqref{Stat} - \eqref{dual} act as a set of necessary and sufficient conditions that the optimal value shall satisfy. To incorporate the KKT conditions in the NN training (and by that also render it a PINN), we denote the discrepancies from zero in \eqref{Stat} - \eqref{dual} with $\epsilon$, as shown in \eqref{EStat}-\eqref{Primal}, and minimize $\epsilon$ as part of the NN loss function, as shown in \eqref{MAE}.
The proposed PINN structure is given in Fig. \ref{PINN_Fig}. The dual variables required for calculating the discrepancy in the KKT conditions are predicted using a separate set of hidden layers. 

The discrepancy in KKT conditions are calculated as follows: 
\begin{align}
&\begin{aligned}
    \epsilon_{stat} &= \lvert \mathbf{c} + \hat{\lambda} + \mathbf{\hat{\overline{\mu}}_g} - \mathbf{\hat{\underline{\mu}}_g} + \mathbf{\hat{\overline{\mu}}_{l}PTDF} - \mathbf{\hat{\underline{\mu}}_{l}PTDF}\lvert
    \label{EStat}
\end{aligned}
\end{align}
\begin{align}
&\begin{aligned}
    \epsilon_{comp} &= \sum_{i=1}^{N_{g}} \left[\lvert  {\hat{\overline{\mu}}_{g,i} \left(P_{g,i}^{max} - \hat{P}_{g,i}\right)} \lvert + \lvert  {\hat{\underline{\mu}}_{g,i}\left(\hat{P}_{g,i} - P_{g,i}^{min}\right)}\lvert \right]\\
   &\qquad + \sum_{j=1}^{N_{l}} \left[ \lvert\hat{\overline{\mu}}_{l,j} \left( \mathbf{PTDF_j(\hat{P}_{g} - P_{d})} - \mathbf{P_{l,j}^{max}} \right) \lvert  \right] \\
    &\qquad +\sum_{j=1}^{N_{l}} \left[  \lvert  \hat{\overline{\mu}}_{l,j} \left( \mathbf{-PTDF_j(\hat{P}_{g} - P_{d})} - \mathbf{P_{l,j}^{max}} \right) \lvert \right]
\end{aligned}\label{EComSlak}\\
&\begin{aligned}
    \epsilon_{dual} &= 
    \pi(\hat{\overline{\mu}}_g)+\pi(\hat{\underline{\mu}}_g)+\pi(\hat{\overline{\mu}}_{l})+\pi(\hat{\underline{\mu}}_{l})
    \label{Edual}
\end{aligned}\\
&\begin{aligned}
    \epsilon_{prim} &= \sum_{i=1}^{N_{g}} \left[  \pi{(P_{g,i}^{max} - \hat{P}_{g,i})}  +  \pi{\left(\hat{P}_{g,i} - P_{g,i}^{min}\right)} \right] \\
      &\qquad + \lvert \sum_{i=1}^{N_g} \hat{P}_{g,i} - \sum_{i=1}^{N_d} P_{d,i} \lvert \\
    &\qquad + \sum_{j=1}^{N_{l}}  \pi \left( \mathbf{PTDF_j(\hat{P}_{g} - P_{d})} - {P_{l,j}^{max}} \right) \\
   &\qquad + \sum_{j=1}^{N_{l}}  \pi \left( \mathbf{- PTDF_j(\hat{P}_{g} - P_{d})} - {P_{l,j}^{max}} \right) 
    \end{aligned}
    \label{Primal}
\end{align}
where $\hat{\lambda}$, $\hat{\mu}_g$ and $\hat{\mu}_l$ are the dual variables predicted using the second set of hidden layers, represented by $\mathbf{L_m}$ in Fig.~\ref{PINN_Fig}. The  $\epsilon_{stat}$ is the absolute value of the stationarity condition, and $\epsilon_{com}$ is the sum of all the complementary slackness conditions given in \eqref{ComSlak1}-\eqref{ComSlak4}. The ReLU activation function, represented by $\pi$, is used to measure the constraint violation, $\epsilon_{prim}$, as shown in \eqref{Primal}. If the NN prediction is the optimal value, then these error factors given in \eqref{EStat} - \eqref{Primal} will be zero. 

By including the KKT conditions in the NN loss, we can have a set of collocation points in the training set. The collocation points are a set of random input values from the input domain. However, unlike the training data points, we have not pre-calculated and do not provide the optimal generation dispatch values $\mathbf{P_g}$ or dual variables $\mathbf{L_m}$ associated with them during training. Instead, the discrepancy in the KKT conditions caused by the predicted active power generation will be used to train the NN; i.e. based on \eqref{MAE} the NN during training seeks weights and biases to minimize this discrepancy, see $MAE_{\epsilon}$. Furthermore, as we will see in Section~\ref{sec:results}, the error factor corresponding to the primal conditions given in \eqref{Primal} can be used without the other conditions to penalize only the constraint violations; this will be similar to the method proposed in \cite{Pred_AC_OPF}.  

The shared parameters of the two neural networks are optimized by minimizing the loss function: 
\begin{align}
&\begin{aligned}
    MAE & = \frac{1}{N_t} \sum_{i=1}^{N_t}  \Lambda_P\underbrace{\lvert \hat{P}_{g} - P_{g}\lvert}_{MAE_p} + \Lambda_L\underbrace{ \lvert \hat{L}_{m} - L_{m}\lvert}_{MAE_l} \\
    &\qquad + \frac{\Lambda_\epsilon}{N_t + N_c} \sum_{i=1}^{N_t + N_c} \underbrace{\epsilon_{stat} + \epsilon_{comp} + \epsilon_{dual} + \epsilon_{prim}}_{MAE_\epsilon}
    \label{MAE}
\end{aligned}
\end{align}
where $N_t$ is the number of training data points, and $N_c$ is the number of collocation points. Furthermore, $MAE_p$, $MAE_l$, and $MAE_\epsilon$ are the mean absolute errors corresponding to active power dispatch, dual variables, and KKT condition violations and $\Lambda_P$, $\Lambda_L$, and $\Lambda_\epsilon$ are their corresponding weights. The PINN performance depends significantly on the weights $\Lambda_P$, $\Lambda_L$, and $\Lambda_\epsilon$. So, they have to be selected appropriately to reduce either the average error or the maximum constraint violations. 

For collocation points, since we do not provide the optimal generation dispatch values $\mathbf{P_g}$ or dual variables $\mathbf{L_m}$ associated with them, both $MAE_p$ and $MAE_l$ will be considered zero (in contrast with the points drawn from the training dataset), and $MAE_\epsilon$ will be used to calculate the training error.  

\subsection{Worst Case Guarantees}
This section describes the worst-case guarantees used for evaluating the performance of the PINN. Worst-case guarantees provide an upper bound to constraint violations, sub-optimality, and distance between optimal generation value and the PINN predictions. To determine these worst-case guarantees, the trained NN will be reformulated into a MILP  problem using the method proposed in \cite{Andreas}. After the training is finished, and the NN is ready to be deployed in practice, \eqref{EStat} -\eqref{Primal} are not used. So, we can ignore them during the MILP formulation.  

The NN formulation given in \eqref{NN1} is linear, so we can directly use it in the MILP problem. However, the nonlinear ReLU activation \eqref{Relu} in the NN has to be reformulated into a mixed integer linear problem as follows:
\begin{align}
    Z^i_k &\leq Z^{' i}_k-Z^{min,i}_k (1-y^i_k) \ &\forall k = 1, ...,K \ \forall i = 1, ...,N_k \label{RelU1}\\
    Z^i_k &\geq Z^{' i}_k \ &\forall k = 1, ...,K \ \forall i = 1, ...,N_k \label{RelU2}
\end{align}
\begin{align}
    Z^i_k &\leq Z^{max,i}_k y^i_k \ &\forall k = 1, ...,K \ \forall i = 1, ...,N_k\label{RelU3} \\
    Z^i_k &\geq 0 \ &\forall k = 1, ...,K \ \forall i = 1, ...,N_k \label{RelU4}\\ 
    y_k &\in \{0, 1\}^{N_k} \ &\forall k = 1, ...,K   \label{RelU5} 
\end{align}
where $Z^i_k$ and $Z^{' i}_k$ are the outputs and inputs of the ReLU activation function, $ Z^{min,i}$ and $Z^{max,i}$ are large value so that they won't be binding, and $y^i_k$ is a binary variable. If $Z^{' i}_k$ is less than zero then $y^i_k$ will be zero and \eqref{RelU3} and \eqref{RelU4} will be active and $Z^i_k$  will be constrained to zero. Else, $y^i_k$ will be equal to one and \eqref{RelU1} and \eqref{RelU2} will make sure $Z^i_k$ is equal to  $Z^{' i}_k$.  
\subsubsection{Worst-Case Guarantees for Constraint Violations}
In this section, the MILP problem formulations used to determine the maximum constraint violations in generator active power outputs, denoted by $v_g$, and line flow violations, denoted by $v_l$, as a result of the  physics-informed neural  network predictions are discussed. The maximum constraint violations in generator active power outputs can be formulated as follows:
\begin{align}
    \underset{\mathbf{\hat{P}_g,P_d,Z,Z^{'},y}}{\mathrm{max}} v_g \label{WCeq1} \\
    v_g = \mathrm{max}(\mathbf{\hat{P}_g - P_g^{max}}, \mathbf{P_g^{min} - \hat{P}_g},0) \\
    s.t. \eqref{NN1}, \eqref{RelU1}- \eqref{RelU5}
\end{align}

Please note $v_g$ is not the maximum constraint violation of a single generator, but rather it is the maximum constraint violation considering all the generators for the entire defined input domain. Similarly, $v_l$ can be determined as follows:
\begin{align}
    \underset{\mathbf{\hat{P}_g,P_d,Z,Z^{'},y}}{\mathrm{max}} v_l \\
    v_l = \mathrm{max}(\lvert \mathbf{PTDF(\hat{P}_g - P_{d})} \lvert - {P_{l}^{max}},0) \\
    s.t. \eqref{NN1}, \eqref{RelU1}- \eqref{RelU5} \label{WCeq2}
\end{align}
where $v_l$  is the overall non-negative maximum line flow constraint violation in the entire input domain. When these MILP problems are solved to the zero MILP gap, we can ensure that the $v_g$ and  $v_l$ values we obtain are the global optima. Thus, we can guarantee that there is no input $\mathbf{P_d}$ in the entire input domain, leading to constraint violations larger than the obtained values $v_g$ and $v_l$.

\subsubsection{Worst-Case Guarantees for Distance of Predicted to Optimal Decision Variables and for Sub-Optimality}
This section establishes the MILP formulations used to determine the maximum distance between the physics-informed neural network prediction and the optimal value, denoted by $v_{dist}$, and the maximum sub-optimality, denoted by $v_{opt}$, in the entire input domain. The formulation used to determine $v_{dist}$ is as follows:
\begin{align}
    v_{dist} = \mathrm{max} \left( \frac{\lvert \mathbf{\hat{P}_g - P_g} \lvert}{ \mathbf{P_g^{max} - P_g^{min}}}\right) \label{WCs1}\\
    \underset{\mathbf{\hat{P}_g,{P}_g,P_d,Z,Z^{'},y}}{\mathrm{max}} v_{dist}\\
    s.t. \eqref{NN1} , \eqref{Stat} - \eqref{dual}, \eqref{RelU1} - \eqref{RelU5}
\end{align}
where $\mathbf{P_g}$ is the optimal generation active power output for a given $\mathbf{P_d}$ calculated in the lower-level optimization problem utilizing the KKT formulation given in \eqref{Stat} - \eqref{dual}, and $\mathbf{\hat{P}_g}$ is the NN prediction.  Similarly, the maximum sub-optimality of the PINN prediction can be formulated as follows:
\begin{align}
    v_{opt} = \mathbf{c^T}(\mathbf{\hat{P}_g - P_g})\\
    \underset{\mathbf{\hat{P}_g,{P}_g,P_d,Z,Z^{'},y}}{\mathrm{max}} v_{opt}\\
    s.t. \eqref{NN1} , \eqref{Stat} - \eqref{dual}, \eqref{RelU1} - \eqref{RelU5} 
\end{align}
By maximizing $v_{opt}$ in the objective function, we can compute worst-case guarantees for the sub-optimality of the predicted solution. The complementary slackness conditions, given in \eqref{ComSlak1} - \eqref{ComSlak4}, are non-linear so they have to be reformulated into linear equations using the Fortuny-Amat McCarl linearization \cite{Fortuny} as follows:
\begin{align}
    \mathbf P^{min}_g - \mathbf P_g &\geq -\mathbf r^{min}_{g}\mathbf M^{min}_{g}\\ \overline{\mu}_{g} &\leq (1-\mathbf r^{min}_{g})\mathbf M^{min}_{g}  \\
    \mathbf P_g - \mathbf P^{max}_g &\geq -\mathbf r^{max}_{g}\mathbf M^{max}_{g} \\
    \underline{\mu}_{g} &\leq (1-\mathbf r^{max}_{g})\mathbf M^{max}_{g}\\
    \mathbf P^{min}_{line} - \mathbf{PTDF}(\mathbf P_g - \mathbf P_d) &\geq  -\mathbf r^{min}_{line}\mathbf M^{min}_{line}  \\
    \overline{\mu}_{l}  &\leq (1-\mathbf r^{min}_{line})\mathbf M^{min}_{line} \\
    \mathbf{PTDF}(\mathbf P_g - \mathbf P_d) - \mathbf P^{max}_{line} &\geq -\mathbf r^{max}_{line}\mathbf M^{max}_{line} \\
    \underline{\mu}_{l}  &\leq  (1-\mathbf r^{max}_{line})\mathbf M^{max}_{line}\label{WCs2}
\end{align}
where $\mathbf r$ is a binary variable and $\mathbf M$ is a sizeable non-binding constant for each condition. The constant M has to be chosen sufficiently large for it to be nonbinding. When the resulting MILP optimization problem is solved to zero MILP gap, we obtain the provable guarantee that there is no input $\mathbf{P_d}$ in the entire input domain that will result in a PINN output with distance or sub-optimality larger than the obtained values of $v_{dist}$ and $v_{opt}$.
\section{RESULTS \& DISCUSSION}
\label{sec:results}
\subsection{Simulation Setup}
\begin{table}[]
\centering
\caption{TEST CASE CHARACTERISTICS}
\begin{tabular}{ccccccc}
\hline \hline
Test cases & $N_{bus}$  & $N_{d}$  & $N_{g}$ & $N_{line}$ & \begin{tabular}[c]{@{}c@{}}Max. loading\\ MW\end{tabular} \\ \hline \hline
case39     & 39  & 21  & 10  & 46    & 6254                                                      \\ \hline
case118    & 118 & 99  & 19 & 186   & 4242                                                      \\ \hline
case162    & 162 & 113 & 12 & 284   & 7239                                                      \\ \hline \hline
\end{tabular}
\label{TC}
\end{table}
We evaluated the effectiveness of PINNs on three PGLib-OPF networks v19.05 \cite{PGLib}. The test case specifications are given in Table \ref{TC}. In each of these test cases, the input domain for each active power demand is assumed to be between 60\% to 100\% of its maximum loading.  The maximum loading was defined according to \cite{PGLib}, and the sum of maximum loading is given in Table I. We used Latin hypercube sampling \cite{hypercube} to randomly generate 100,000 samples from the input domain. Of the 100,000 samples generated, 20\% was used as training and test dataset points (i.e we calculated and associated with them the DC-OPF results), 50\% was used as collocation points, and the rest was used to determine average errors in an unseen test set. For the data points in training and test sets, the MATPOWER DC-OPF solver \cite{MATPOWER} was used to determine the optimal active power generation. Then the KKT Conditions, given in \eqref{EStat} -\eqref{Edual}, were utilized to determine the values of Lagrange multipliers.   

The NN architecture consists of two sets of hidden layers, as shown in Fig \ref{PINN_Fig}.  The NN used to predict the optimal active power generations has three hidden layers with 20 neurons each; the NN used to obtain the Lagrange multipliers has three hidden layers with 30 neurons each. We used TensorFlow  \cite{tensorflow} for NN training, we fixed the maximum number of training epochs to 5'000, and split the data set into two batches. The mean absolute error was used to determine the loss between NN predictions and the actual optimal solution during training. 

The MILP problem used for worst-case guarantees was formulated in YALMIP \cite{yalmip} and solved using Gurobi. After solving the MILPs, we verified that the complementary slackness conditions are satisfied, and the constants are non-binding. A laptop with AMD Ryzen 7 pro CPU, 16 GB RAM, and Radeon GPU was used to carry out the computational experiments. The code to reproduce all simulation results is available online \cite{PINN_Code}. 
\subsection{Physics-Informed Neural Network Average Performance over Test Set Samples}
In the following, we evaluate the average performance of four different configurations of the PINN and compare them to a standard NN without the PINN layer, specified as NN in Table~\ref{TableAvg}. The first PINN configuration, represented by  Pg Abs, includes the absolute value of the generation limit violation term given in \eqref{Primal}, while all other KKT condition violations, given in \eqref{EStat} - \eqref{Primal}, are ignored. We do this to understand how penalizing the generation constraint violation impacts the overall system performance. Similarly, the Pg Square and Pg Exp have the generation limit violation term given in \eqref{Primal} in square and exponential terms. Finally, in KKT, all the KKT constraint violations given in \eqref{EStat} - \eqref{Primal}  are included in the PINN loss term.  

The metrics used for comparing the average performance in Table~\ref{TableAvg} are:
\begin{enumerate}
    \item Mean absolute error (MAE) in percentage.
    \item Average generation active power constraint violation $v_g$ in MW.
    \item Average line flow limit violations $v_l$ in MW.
    \item Average distance of predicted value to optimal decision variables $v_{dist}$ in percentage.
    \item Average sub-optimality $v_{opt}$ in percentage.
\end{enumerate}

\begin{table}[]

\centering
\caption{Performance Averaged Over Test Set Samples}
\resizebox{\columnwidth}{!}{%
\begin{tabular}{cccccccc}
\hline \hline
\multicolumn{3}{c}{\multirow{2}{*}{TestCase}}                & \multirow{2}{*}{\begin{tabular}[c]{@{}c@{}}MAE\\      (\%)\end{tabular}} & \multirow{2}{*}{\begin{tabular}[c]{@{}c@{}}$v_g$\\      (MW)\end{tabular}} & \multirow{2}{*}{\begin{tabular}[c]{@{}c@{}}$v_{line}$\\      (MW)\end{tabular}} & \multirow{2}{*}{\begin{tabular}[c]{@{}c@{}}$v_{dist}$ \\      (\%)\end{tabular}} & \multirow{2}{*}{\begin{tabular}[c]{@{}c@{}}$v_{opt}$\\      (\%)\end{tabular}} \\
\multicolumn{3}{c}{}                                         &                                                                          &                                                                         &                                                                            &                                                                             &                                                                           \\ \hline \hline
\multirow{5}{*}{Case 39}  & \multicolumn{2}{c}{NN}           & 0.23                                                                     & 2.58                                                                    & 0.00                                                                       & 0.35                                                                        & 0.15                                                                      \\ \cline{2-8} 
                          & \multirow{4}{*}{PINN} & Pg   Abs & 0.05                                                                     & 0.44                                                                    & 0.01                                                                       & 0.13                                                                        & 0.02                                                                      \\ \cline{3-8} 
                          &                       & Pg Sqr   & 0.24                                                                     & 2.13                                                                    & 0.00                                                                       & 0.59                                                                        & 0.12                                                                      \\ \cline{3-8} 
                          &                       & Pg Exp   & 0.10                                                                     & 0.90                                                                    & 0.02                                                                       & 0.15                                                                        & 0.07                                                                      \\ \cline{3-8} 
                          &                       & KKT      & 0.12                                                                     & 0.92                                                                    & 0.04                                                                       & 0.39                                                                        & 0.01                                                                      \\ \hline
\multirow{5}{*}{Case 118} & \multicolumn{2}{c}{NN}           & 0.68                                                                     & 0.66                                                                    & 2.75                                                                       & 8.42                                                                        & 0.65                                                                      \\ \cline{2-8} 
                          & \multirow{4}{*}{PINN} & Pg   Abs & 0.89                                                                     & 1.57                                                                    & 5.87                                                                       & 9.49                                                                        & 0.74                                                                      \\ \cline{3-8} 
                          &                       & Pg Sqr   & 0.69                                                                     & 0.58                                                                    & 2.54                                                                       & 9.20                                                                        & 0.53                                                                      \\ \cline{3-8} 
                          &                       & Pg Exp   & 1.01                                                                     & 3.23                                                                    & 5.92                                                                       & 8.70                                                                        & 0.61                                                                      \\ \cline{3-8} 
                          &                       & KKT      & 1.24                                                                     & 3.01                                                                    & 4.60                                                                       & 4.60                                                                        & 8.28                                                                      \\ \hline
\multirow{5}{*}{Case 162} & \multicolumn{2}{c}{NN}           & 3.48                                                                     & 8.65                                                                    & 11.54                                                                      & 23.70                                                                       & 0.45                                                                      \\ \cline{2-8} 
                          & \multirow{4}{*}{PINN} & Pg   Abs & 3.35                                                                     & 10.31                                                                   & 11.03                                                                      & 23.54                                                                       & 0.62                                                                      \\ \cline{3-8} 
                          &                       & Pg Sqr   & 3.43                                                                     & 9.05                                                                    & 11.25                                                                      & 23.77                                                                       & 0.53                                                                      \\ \cline{3-8} 
                          &                       & Pg Exp   & 3.30                                                                     & 1.12                                                                    & 7.09                                                                       & 24.04                                                                       & 0.24                                                                      \\ \cline{3-8} 
                          &                       & KKT      & 3.11                                                                     & 5.34                                                                    & 7.54                                                                       & 22.31                                                                       & 0.49                                                                      \\ \hline \hline
\end{tabular}}
\label{TableAvg}
\end{table}
During the analysis, we observed that both the average and the worst-case performance of the PINN depends a lot on the hyper-parameter values, i.e., $\Lambda_P$, $\Lambda_L$, and $\Lambda_\epsilon$ weights. We experimented with different hyper-parameter values, and the ones which offered the lowest worst-case generation constraint violation are used to produce the results given in Table \ref{TableAvg}. Because of this, the average performance compared to the standard NN has worsened in some cases, especially in the case of the 162 bus system compared to other sets of results we obtained. This could be because of the limited number of hyper-parameter values we explored. Even then, in most cases, the mean absolute error and the average constraint violations and sub-optimality have either improved or remain comparable. This indicates the satisfactory generalization capability of the PINN. It shall also be noted that due to additional number of equations and collocation points, the PINN was observed to take almost thrice as much time to train as opposed to the standard NN.

\subsection{Worst-Case Guarantees for Constraint Violations}
Using the mixed-integer linear reformulation given in \eqref{WCeq1}-\eqref{WCeq2}, we solve the MILPs to compute the corresponding worst-case guarantees. The results are shown in Table \ref{TableWC}. As hypothesized, when the absolute values of the generation limit violation were added to the NN loss, the worst-case generation constraint violation was reduced by at least 25\%. Moreover, when we used higher-order terms to estimate the loss, the worst-case generation constraint violation values were reduced even further, and by at least 20\% in all cases. This indicates that we can achieve a better worst-case guarantee by using higher-order terms to penalize constraint violations. When we added all the KKT conditions violations to the loss function (all in absolute terms), the worst-case generation and line flow constraint violation were reduced further in the 39-bus and 118-bus systems. In the 162-bus system, the results are comparable to that of the Pg Exp. This validates the hypothesis that we can achieve a better worst-case guarantee by incorporating KKT condition violations into the NN training.  
\begin{table}[]
\centering
\caption{Worst-Case Guarantees for Constraint Violations}
\begin{tabular}{ccccccc}
\hline \hline
\multicolumn{3}{c}{\multirow{2}{*}{Test Cases}}            & \multicolumn{2}{c}{{$v_g$}}                                                     & \multicolumn{2}{c}{{$v_{line}$}}                                                  \\ \cline{4-7} 
\multicolumn{3}{c}{}                                       & {MW} & {\begin{tabular}[c]{@{}c@{}}\%   wrt \\ max load\end{tabular}} & {MW} & {\begin{tabular}[c]{@{}c@{}}\%   wrt \\ max load\end{tabular}} \\ \hline \hline
\multirow{5}{*}{Case 39}  & \multicolumn{2}{c}{NN}         & 365         & 6                                                                     & 146         & 2                                                                     \\ \cline{2-7} 
                          & \multirow{4}{*}{PINN} & Pg Abs & 265         & 4                                                                     & 63          & 1                                                                     \\ \cline{3-7} 
                          &                       & Pg Sqr & 195         & 3                                                                     & 68          & 1                                                                     \\ \cline{3-7} 
                          &                       & Pg Exp & 144         & 2                                                                     & 87          & 1                                                                     \\ \cline{3-7} 
                          &                       & KKT    & 133         & 2                                                                     & 54          & 1                                                                     \\ \hline
\multirow{5}{*}{Case 118} & \multicolumn{2}{c}{NN}         & 572         & 13                                                                    & 252         & 6                                                                     \\ \cline{2-7} 
                          & \multirow{4}{*}{PINN} & Pg Abs & 246         & 6                                                                     & 131         & 3                                                                     \\ \cline{3-7} 
                          &                       & Pg Sqr & 209         & 5                                                                     & 94          & 2                                                                     \\ \cline{3-7} 
                          &                       & Pg Exp & 177         & 4                                                                     & 114         & 3                                                                     \\ \cline{3-7} 
                          &                       & KKT    & 209         & 5                                                                     & 95          & 2                                                                     \\ \hline
\multirow{5}{*}{Case 162} & \multicolumn{2}{c}{NN}         & 1725        & 24                                                                    & 1224        & 17                                                                    \\ \cline{2-7} 
                          & \multirow{4}{*}{PINN} & Pg Abs & 861         & 12                                                                    & 1193        & 16                                                                    \\ \cline{3-7} 
                          &                       & Pg Sqr & 792         & 11                                                                    & 1209        & 17                                                                    \\ \cline{3-7} 
                          &                       & Pg Exp & 672         & 9                                                                     & 657         & 9                                                                     \\ \cline{3-7} 
                          &                       & KKT    & 696         & 10                                                                    & 475         & 7                                                                     \\ \hline \hline
\end{tabular}
\label{TableWC}
\end{table}
\subsection{Worst-Case Guarantees for (i) Distance of Predicted to Optimal Decision Variables and (ii) for Sub-Optimality}
The MILP formulation given in \eqref{WCs1} - \eqref{WCs2} was used to find the worst-case guarantees for the distance of predicted to optimal decision variables and sub-optimality. The results given in Table \ref{TableWS} are calculated for the same set of hyperparameter values used to obtain the results presented in Table \ref{TableWC}. In some cases, adding only the constraint violation terms to the loss function has a negative effect on the solution's optimality. Results still show, however, that adding all the KKT condition violations in the loss term obtains the best performance, as the maximum distance between the PINN prediction and the optimal value, as well as the worst-case suboptimality have improved in all three cases.
\begin{table}[]
\centering
\caption{Worst-Case Guarantees For (i) Distance Of Predicted to Optimal Decision Variables and (ii) Suboptimality}
\begin{tabular}{ccccc}
\hline \hline
\multicolumn{3}{c}{\multirow{2}{*}{Test Cases}}              & {$v_{dist}$} & {$v_{opt}$} \\ \cline{4-5} 
\multicolumn{3}{c}{}                                         & {\%}    & {\%}   \\ \hline \hline
\multirow{5}{*}{Case 39}  & \multicolumn{2}{c}{NN}           & 78             & 10            \\ \cline{2-5} 
                          & \multirow{4}{*}{PINN} & Pg   Abs & 71             & 10            \\ \cline{3-5} 
                          &                       & Pg Sqr   & 72             & 6             \\ \cline{3-5} 
                          &                       & Pg Exp   & 86             & 11            \\ \cline{3-5} 
                          &                       & KKT      & 69             & 9             \\ \hline 
\multirow{5}{*}{Case 118} & \multicolumn{2}{c}{NN}           & 542            & 13            \\ \cline{2-5} 
                          & \multirow{4}{*}{PINN} & Pg   Abs & 465            & 9             \\ \cline{3-5} 
                          &                       & Pg Sqr   & 382            & 9             \\ \cline{3-5} 
                          &                       & Pg Exp   & 658            & 17            \\ \cline{3-5} 
                          &                       & KKT      & 306            & 7             \\ \hline 
\multirow{5}{*}{Case 162} & \multicolumn{2}{c}{NN}           & 788            & 58            \\ \cline{2-5} 
                          & \multirow{4}{*}{PINN} & Pg   Abs & 488            & 65            \\ \cline{3-5} 
                          &                       & Pg Sqr   & 838            & 45            \\ \cline{3-5} 
                          &                       & Pg Exp   & 469            & 65            \\ \cline{3-5} 
                          &                       & KKT      & 482            & 69            \\ \hline \hline
\end{tabular}
\label{TableWS}
\end{table}
\section{Conclusion and Future Work}
This paper presents two key contributions. First, to the best of our knowledge, this is the first paper to propose physics-informed neural networks for optimal power flow applications. 
We show that by combining the KKT conditions with the neural network, the physics-informed neural network achieves higher accuracy while utilizing substantially fewer data points. Second, we extend our previous work on worst-case guarantees to cover the physics-informed neural networks (PINNs), and we show that PINNs result in lower worst-case violations than conventional neural networks. Future work includes the extension of the proposed approaches to AC-OPF problems and a multilevel optimization algorithm to determine the key PINN hyper-parameters that improve average and worst-case performance.



\markeverypar{\the\everypar\looseness=0 } 
\bibliographystyle{IEEEtran}
\bibliography{R1}
\end{document}